\documentclass[epj]{svjour}
\usepackage{amssymb,amsmath}
\usepackage{epsfig}
\usepackage{cite}

\newcommand{\qs}{Q_{\rm s, p}}
\newcommand{\qsh}{Q_{\rm s, h}}
\newcommand{\qsa}{Q_{\rm s, A}}

\begin{document}
\title{Nuclear size and rapidity dependence of the saturation
scale from QCD evolution and experimental data
\thanks{Joint contribution to the proceedings of  the Hard Probes 2004
Conference based on the oral presentations by J.G. Mi\-lhano and C.A. Salgado}
}
%\subtitle{A study of the non-linear Balitsky-Kovchegov evolution and experimental data}
\author{Javier L. Albacete\inst{1},  N\'estor Armesto\inst{2,} \inst{3},
J. Guilherme Milhano \inst{2,} \inst{4},  Carlos A. Salgado\inst{2} 
and Urs Achim Wiedemann \inst{2}% etc
}                     % Do not remove
\institute{Departamento de F{\'\i}sica, Universidad de C\'ordoba, E-14071
C\'ordoba, Spain
\and 
Department of Physics, CERN, Theory Division, CH-1211 Gen\`eve,
Switzerland  
\and
Departamento de F{\'\i}sica de Part{\'\i}culas, Universidade de Santiago de
Compostela, E-15706 Santiago de Compostela, Spain 
\and 
CENTRA, Instituto Superior T\'ecnico (IST), Av. Rovisco Pais 1, P-1049-001
Lisboa, Portugal}
\date{Received: date / Revised version: date}
\abstract{
The solutions of the Balitsky-Kovchegov evolution equations are studied
numerically and compared with known analytical estimations. The rapidity and
nuclear size dependences of the saturation scale are obtained for the cases of
fixed and running coupling constant. These same dependences are studied in
experimental data, on lepton-nucleus, deuteron-nucleus and nucleus-nucleus
collisions, through geometric scaling and compared with the theoretical
calculations.
 % end of PACS codes
} %end of abstract
\titlerunning{Nuclear size and rapidity dependence of the saturation scale}
\authorrunning{Albacete \textit{et al.}}

\maketitle
\section{Introduction}
\label{intro}
The cleanest experimental information about parton distributions comes
 from deep inelastic scattering experiments.
%The cleanest system where parton distributions can be studied is in
%lepton-hadron deep inelastic scattering experiments. 
At high energy, they can be
described by the QCD dipole model~\cite{dipole}, which expresses the cross
section of the virtual photon, emitted by the lepton, on the hadron $h$
(proton or nucleus) as
\begin{eqnarray}
  \sigma_{T,L}^{\gamma^* h}(x,Q^2) = \int d{\bf r} \hspace{-0.1cm}
  \int_0^1
  \hspace{-0.2cm} dz
  \vert \Psi_{T,L}^{\gamma^*}(Q^2,{\bf r},z)\vert^2\,
  \sigma_{\rm dip}^h({\bf r},x)\, .
  \label{eq1}
\end{eqnarray}
Here, $\Psi_{T,L}$ are the %perturbatively computed 
transverse and 
longitudinal wave functions for the splitting of $\gamma^*$ into 
a $q\bar{q}$ dipole of transverse size ${\bf r}$ with light-cone 
fractions $z$ and $(1-z)$. % carried by the quark and antiquark respectively.
The dipole cross section,
$\sigma_{\rm dip}^h({\bf r},x)$, can be written as an integral of 
the dipole scattering amplitude $N_h$ over the impact parameter 
${\bf b}$,
\begin{eqnarray}
  \sigma_{\rm dip}^h({\bf r},x) = 2 \int d{\bf b}\,
                                     N_h({\bf r},x;{\bf b})\, .
  \label{eq2}
\end{eqnarray}
In this framework, the QCD evolution is included in the dipole forward 
scattering
amplitude. The simplest equation describing this evolution and taking into
account saturation effects is the Balitsky--Kovchegov (BK) 
equation \cite{bkeq}. %In this contribution 
We will first review the properties 
of the solutions of the BK equations in configuration space and then discuss
whether these properties are observed in presently available  data.  

\section{BK equation}
\label{sec:bk}

The BK equation \cite{bkeq} describes 
the rapidity $Y=\ln (s/s_0) = \ln (x_0/x)$ evolution of the scattering 
probability $N(\vec{x},\vec{y},Y)$ of a $q\bar q$ dipole with an hadronic target. When considering an  homogeneous target with radius much larger than the size of any dipole, the dependence on impact parameter can be neglected and the equation reads
\begin{multline}
\frac{\partial N(r,Y)}{\partial Y}=\int {d^2z\over 2\pi}\, K(\vec{r},
\vec{r}_1,\vec{r}_2)
\Big[N(r_1,Y)\\
+N(r_2,Y)-N(r,Y)-N(r_1,Y)N(r_2,Y)\Big],
\label{eq:bk}
\end{multline}
where we define $\vec{r}=\vec{x}-\vec{y},\, \vec{r}_1=\vec{x}-\vec{z},\, 
\vec{r}_2=\vec{y}-\vec{z}.$ 
%\begin{equation}
%\vec{r}=\vec{x}-\vec{y},\ \ 
%\vec{r}_1=\vec{x}-\vec{z},\ \ 
%\vec{r}_2=\vec{y}-\vec{z},
%\label{eq:coord}
%\end{equation}
Here, $\vec{x}$ ($\vec{y}$) is the position of the $q$ ($\bar q$) in 
transverse space with respect to the centre of the target and $\vec{z}$ is the 
corresponding one for the emitted gluon. The BFKL kernel is given by 
\begin{equation}
K(\vec{r},
\vec{r}_1,\vec{r}_2)=\bar \alpha_s\,{r^2 \over r_1^2r_2^2}\ \ ,
\ \ \bar \alpha_s={\alpha_sN_c\over \pi}\,.
\label{eq:kernelfc}
\end{equation}

The BK equation (\ref{eq:bk}) has a rather simple probabilistic interpretation \cite{Kovner:2000pt}:
when evolved in rapidity, the parent dipole with ends located at $\vec{x}$ and $\vec{y}$ emits a gluon, which, in the large-$N_c$ limit, corresponds to two dipoles with ends $(\vec{x},\vec{z})$ and $(\vec{z},\vec{y})$, respectively. 
The probability of such emission is given by the BFKL kernel (\ref{eq:kernelfc}), and weighted by the scattering probability of the new dipoles minus the scattering probability of the parent dipole (as the
variation with rapidity of the latter is computed). 
The non-linear term, subtracted in order to avoid double counting, prevents, in contrast to BFKL, the amplitude from growing boundlessly with rapidity. 
The BK equation ensures unitarity locally in transverse configuration space, $|N(r,Y)|\leq 1$. This is guaranteed since
for $N(r,Y)=1$, the derivative with respect to $Y$ in (\ref{eq:bk}) cannot be positive.

The BK equation (\ref{eq:bk}) was derived at leading order in $\alpha_s\ln{(s/s_0)}$ for a fixed coupling constant $\alpha_s$.
It is expected that  next-to-leading-log corrections will play a significant role. An important part of these corrections will arise, as in BFKL, from the running of the coupling.
The scale of the running coupling can only be determined in earnest once the full next-to-leading-log calculation is available.
It is not clear a priori which of the three distance scales in the kernel (\ref{eq:kernelfc}) --- an `external' one, the size of the parent dipole; $r$, and two `internal' ones, the sizes of the two newly created dipoles, $r_1$ and $r_2$ --- should drive the running of the coupling.
In order to access the sensitivity of the results to an heuristically introduced running coupling, different  modifications of the kernel (\ref{eq:kernelfc}) were considered in \cite{Albacete:2004gw}.
Essentially, the different physical cases are accounted for by three modifications of  (\ref{eq:kernelfc}): $K1$, where the parent dipole scale $r$ is used to evaluate the running of the coupling; $K2$, where the sizes $r_1$ and $r_2$ of the created dipoles drive the running of the coupling; and $K3$ which further modifies $K2$ by exponentially  weighting the gluon emission vertex, thus 
imposing short range interactions, in order to check for possible sensitivity 
of the results to the Coulomb tails of the kernel. The coupling has been 
allowed to run in
the standard one-loop form with three flavours. It has been frozen in the
infrared to a value $\alpha_s(k=0)\equiv \alpha_0$, see \cite{Albacete:2004gw}
 for details.

In our numerical implementation of BK evolution (for a detailed discussion see \cite{Albacete:2004gw}) we consider three different initial conditions evolved from some fixed value of $x_0$ (in practice, $x_0 \sim 0.01$).

First, we consider an initial condition with the same $r$-dependence at fixed $x_0$ as the Golec-Biernat--W\"usthoff model (GBW) \cite{Golec-Biernat:1998js}, albeit with the $x$-dependence given by the BK evolution \footnote{Here and in the other initial conditions 
(\ref{eq:mv}), (\ref{eq:as}) below, we denote by $Q_s^{\prime }$ what is
usually called the saturation scale.
Our definition of the saturation scale $Q_s$ is somewhat
different, see Equation (\ref{defqs}) below,
but the relation between both scales is
straightforward e.g. in GBW, $Q_s^{\prime 2}=-4\ln{(1-\kappa)}\, Q_s^2$.}
\begin{equation}
N^{GBW}(r)=1-\exp{\left[-\frac{r^2Q_s^{\prime 2}}{4}\right]}\,.
\label{eq:gbw}
\end{equation}
The second initial condition takes the form given by the
McLerran-Venugopalan model \cite{mvmodel} (MV):
\begin{equation}
N^{MV}(r)=1-\exp{\left[-\frac{r^2Q_s^{\prime 2}}{4}
\ln{\left(\frac{1}{r^2\Lambda_{QCD}^2}+e\right)}\right]}\, .
\label{eq:mv}
\end{equation}
For transverse momenta $k \sim 1/r \geq
{\cal O}(1\, {\rm GeV})$, the sensitivity to the infrared cut off $e$ is 
negligible. The amplitudes $N^{GBW}$ and $N^{MV}$ are similar for
momenta of order $Q_s^\prime$ but differ strongly in their high-$k$
behaviour. The corresponding unintegrated gluon distribution
$\phi(k) = \int \frac{d^2r}{2\pi r^2} e^{i \vec{r}\cdot \vec{k}}\, N(r)$
decays exponentially for $N^{GBW}$ but has a power-law tail
$\sim 1/k^2$ for $N^{MV}$. 
Finally, we consider
\begin{equation}
N^{AS}(r)=1-\exp{\left[-(r\,Q_s^\prime)^c\right]}\, .
\label{eq:as}
\end{equation}
The interest in this ansatz is that the small-$r$ behaviour
$N^{AS} \propto r^c$ corresponds to an anomalous
dimension $1-\gamma = 1- c/2$ of the unintegrated gluon distribution
at large transverse momentum. This anomalous dimension can be chosen
to differ significantly from that of the initial conditions $N^{GBW}$ 
and $N^{MV}$. Our choices $c=1.17$ and $c=0.84$ are somewhat arbitrary.
They can be motivated a posteriori by the observation that 
the anomalous dimension of the evolved BK solution 
for both fixed and running coupling 
lies between the anomalous dimension of the initial conditions 
$N^{AS}$ and $N^{GBW}$ (or $N^{MV}$). 
Thus, the choice of $N^{AS}$ is very convenient to
establish generic properties of the solution of
the  BK equation.

\section{Numerical results}

\subsection{Evolution}

\begin{figure}[h!]
\begin{center}
\epsfig{file=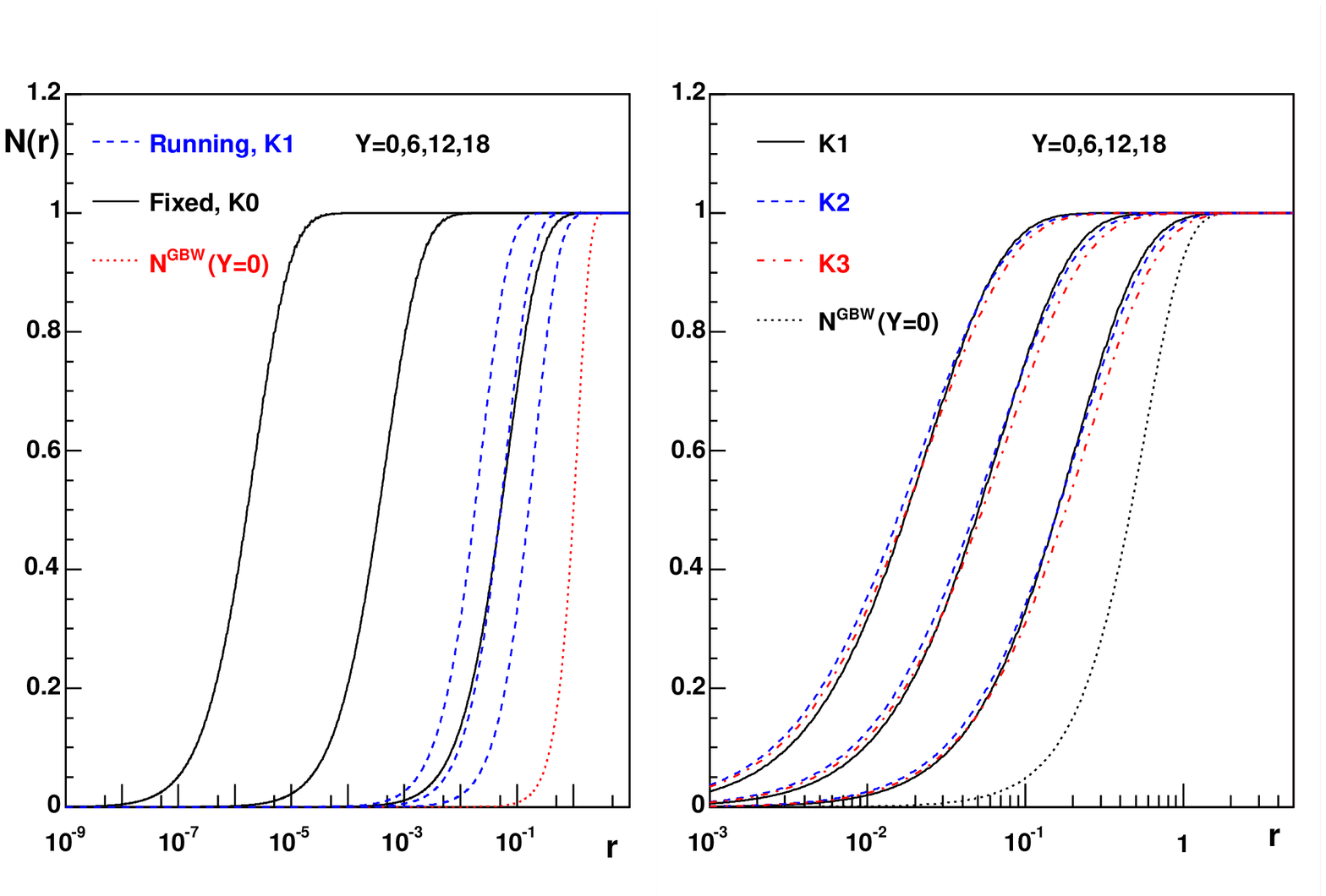,width=0.45\textwidth}
\end{center}
%\vskip -1.cm
\caption{Solutions of the BK equation for GBW initial condition (dotted line)
for rapidities $Y=6$, 12 and 18 with $\bar \alpha_0=0.4$.  Left plot:
Evolution with fixed ($K0$, solid lines) and running coupling ($K1$, dashed
lines). Right plot: evolution with running coupling for kernel modifications
$K1$ (solid lines), $K2$ (dashed lines) and $K3$ (dashed-dotted lines).}
\label{fig:evol}
\end{figure}

Figure \ref{fig:evol} shows the evolution of the dipole scattering 
probability for GBW initial condition with fixed and running 
coupling. 
The evolution is much faster for fixed coupling than 
for running coupling.
Further, the differences arising from the specific implementation of 
running coupling effects 
in the modified BFKL kernels $K1$, $K2$ and $K3$ are rather small and well understood qualitatively  \cite{Albacete:2004gw}.
Importantly, effects of imposing 
short range interactions, $K3$, are very small (unless the range of 
the interaction is unphysically small).

\subsection{Geometrical scaling}

In the limit $Y \to \infty$, the solutions of the BK evolution are no 
longer functions of the variables $r$ and $Y$ separately, but instead 
depend on a single scaling variable
\begin{equation}
\tau\equiv r\,Q_s(Y)\, .
\end{equation}
The saturation momentum $Q_s(Y)$ determines the transverse
momentum below which the unintegrated gluon distribution is saturated.
It can be characterized
by the position of the falloff in $N(r)$,
e.g. via the 
definition
\begin{equation}
  N(r=1/Q_s(Y),Y)=\kappa,
\label{defqs}
\end{equation}
where $\kappa$ is a constant which is smaller than, but of order, one.
Different choices such as $\kappa=1/2$ (as in the results given below)  and $\kappa=1/e$
lead to negligible differences in the determination of $Q_s(Y)$. 

The accuracy of scaling at small $r$ can be checked by comparison with the analytical scaling forms, for fixed and running coupling respectively,  proposed in \cite{Mueller:2002zm}:
\begin{equation}
f^{1)}(\tau)=a\tau^{2\gamma}\left(\ln{\tau^2}+\delta\right)\, ,
\label{eq:f1}
\end{equation}
\begin{equation}
f^{2)}(\tau)=a\tau^{2\gamma}\left(\ln{\tau^2}+{1\over \gamma}\right)\, .
\label{eq:f2}
\end{equation}
Here, $1-\gamma$ is usually called the anomalous dimension which governs 
the leading large-$k$ behaviour of the unintegrated gluon distribution.

We determine $\gamma$ from a fit of our numerical results to the functions
(\ref{eq:f1}) and
(\ref{eq:f2}) in the $Y$-independent region
$10^{-5}<\tau<10^{-1}$, i.e. for $10^5\,Q_s>1/r>10\,Q_s$,
with $a$, $\gamma$ and $\delta$ as free parameters.
The results given below were found to be insensitive to a variation of
the lower limit of this fitting range.

\begin{figure}[t!]
\begin{center}
\epsfig{file=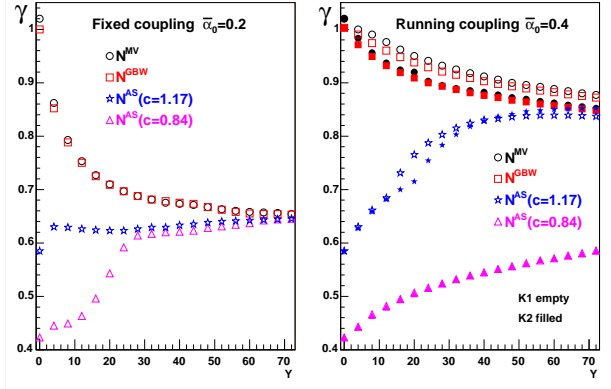,width=0.45\textwidth}
\end{center}
%\vskip -1.cm
\caption{The rapidity dependence of the parameter $\gamma$, characterizing
the anomalous dimension $1-\gamma$, as determined by a fit of  
(\ref{eq:f1}) to the BK
solutions for different initial conditions: GBW (squares),
MV (circles), and AS with $c=1.17$ (stars) and $c=0.84$ (triangles).  Left
plot: results for fixed coupling with $\bar \alpha_0=0.2$. Right plot: results
for running coupling with $\bar \alpha_0=0.4$ and two versions of the kernel
$K1$ (empty symbols) and $K2$ (filled symbols).}
\label{fig:anom}
\end{figure}

For the case of fixed coupling constant, we find that the function
$f^{1)}$ provides a very good fit to the evolved solutions.
In Figure \ref{fig:anom}, we show the fit values of the parameter
$\gamma$, obtained for fixed coupling constant from the
evolution of different initial conditions $N^{GBW}$, $N^{MV}$, and $N^{AS}$
for different values of $c$.
At initial rapidity, these distributions have widely different anomalous
dimensions but evolution drives them to a common
value, $\gamma\simeq 0.65$, which lies close to the theoretically 
conjectured one \cite{Iancu:2002tr,Mueller:2002zm} of $0.628$.
For a small fixed coupling constant $\bar{\alpha}_0=0.2$, this asymptotic
behaviour is reached at $Y \sim 70$, while for a larger coupling constant
$\bar{\alpha}_0=0.4$ the approach to this asymptotic value takes 
half the length of evolution (results not shown). For
fixed coupling solutions, $f^{2)}$ does not provide a good fit to our
numerical results.

%%%%%%%%%%%%%%%%%%%%%%%%%%%%%%%%%%%%%%%%%%%%%%%
%%%%%%%%%%%%%%%%%%%%%%%%%%%%%%%%%%%%%%%%%%%%%%%%%%%%%%%%%%%%%%

We have repeated this comparison for all running coupling prescriptions.
We found that both $f^{1)}$ and $f^{2)}$ provide good fits and yield
very similar values of $\gamma$. The results for $K3$ are numerically
indistinguishable from those for $K2$ and will not be shown in what
follows. Also, the value of $\gamma$ was found to be independent
of the coupling constant $\bar \alpha_0$ at $r\to \infty$.
In Figure \ref{fig:anom}, we show the values of $\gamma$ extracted 
from a fit to $f^{1)}$. Irrespective of the initial condition, they
approach a common asymptotic value $\gamma\sim 0.85$. While our
numerical findings for $N^{AS}$ with $c=0.84$ are not inconsistent
with the approach to this asymptotic value, no firm conclusions can
be drawn. This initial condition just starts too far
away from the asymptotic scaling solution to reach it within the 
numerically accessible rapidity range. In this case, the monotonic 
increase of $\gamma$ with rapidity at large $Y$ is smaller than the 
increase for $N^{AS}$ with $c=1.17$ at comparable values of $\gamma$,
indicating that the rapidity evolution of the anomalous dimension
depends in general not only on the small-$r$ behaviour, but
on the full shape of the scattering probability. 
 
The value $\gamma\sim 0.85$ is considerably larger
than the one found in fixed coupling evolution. This is
in agreement with previous numerical results \cite{Braun:2003un} but
in contrast to theoretical 
expectations \cite{Iancu:2002tr,Mueller:2002zm,Triantafyllopoulos:2002nz} 
which predict the same
value of $\gamma$ for the fixed and running coupling cases.
As an additional check, we have performed running coupling evolution from an
initial condition given by the solution at large
rapidity of fixed coupling evolution
(for which $\gamma \simeq 0.65$). We find that even with this
initial condition,
running coupling
evolution leads to a value of $\gamma \sim 0.85$.

\begin{figure}[t!]
\begin{center}
\epsfig{file=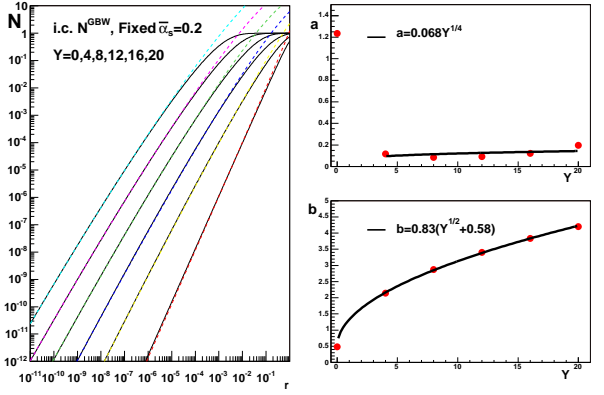,width=0.45\textwidth}
\end{center}
%\vskip -1.cm
\caption{Plot on the left: solutions of the BK equation (solid lines) with GBW
initial condition and fixed coupling $\bar \alpha_s=0.2$ compared to fits
(dashed lines) to the DLL expression (\ref{eq:dll}),
for rapidities $Y=0$, 4, 8, 12, 16 and
20 (curves from right to left). Plots on the right: values of the coefficients
$a(Y)$ and $b(Y)$ (circles)
in the DLL expression versus $Y$, compared to fits (curves) to the
functional form suggested by DLL.}
\label{fig:dll}
\end{figure}

It has been
argued \cite{Iancu:2002tr,Mueller:2002zm} that expressions (\ref{eq:f1}) and
(\ref{eq:f2}) are only
valid for values of $\tau$ inside the scaling window,
$\tau_{\rm
sw}\sim \Lambda_{QCD}
/Q_s(Y)<\tau\lesssim  1$ with $Y_0$ the initial rapidity,
and that the dipole
scattering probability returns to the double-leading-log (DLL) expression
\begin{multline}
N^{DLL}(r)=a(Y)
\,r^2\,[-\ln{(r^2\Lambda^2)}]^{-3/4}\times\\ \times
\exp{\left[b(Y)\sqrt{-\ln{(r^2\Lambda^2)}}
\,\right]}\,,
\label{eq:dll}
\end{multline}
with $a(Y)\propto Y^{1/4}$ and $b(Y)\propto \sqrt{Y}$,
for values $\tau < \tau_{\rm sw}$.
We have checked that this form provides a good fit
(fit and numerical solution
differ by less
than $\pm 10$\%)
to the fixed coupling
solution of BK
for $\tau<\tau_{\rm sw}=
\Lambda/Q_s(Y)$, $\Lambda\sim 0.2$ GeV,
see Figure \ref{fig:dll}.
Our comparison is limited to rapidities $Y\leq 20$, since the scaling window
starts to extend over the entire numerically accessible $r$-space for $Y>20$.
Up to $Y=20$, the coefficients $a(Y)$ and $b(Y)$ follow the expected
DLL $Y$-behaviour,
see Figure \ref{fig:dll}.
However, the
scaling ansatz $f^{1)}$ provides an equally good fit to the BK solutions 
for $\tau<\tau_{\rm sw}$.
This is the reason why in previous numerical studies  
\cite{Albacete:2003iq} no upper bound for a scaling window was found.
When the solutions of BK are fitted to $f^{1)}$
within the scaling window,
the values of $\gamma$ at $Y=0$ for both initial conditions are $\lesssim 20$\%
smaller than those found when the fit is done
within a fixed $\tau$-window.
But for larger $Y$
the values of $\gamma$ extracted from fits within either the
scaling window or some fixed $\tau$-window approach each other and quickly
coincide.

%%%%%%%%%%%%%%%%%%%%%%%%%%%%%%%%%%%%%%%%%%%%%%%

\subsection{Rapidity dependence of the saturation scale}

In the scaling region and for large rapidity (where $Q_s(Y)\gg \Lambda_{QCD}$), the rapidity dependence of the saturation scale is determined \cite{Iancu:2002tr} by
\begin{equation}
\frac{\partial\ln{\left[Q^2_s(Y)/\Lambda^2\right]}}{\partial Y}=d\,\bar
\alpha_s\, ,
\label{qsy}
\end{equation}
where the numerical value of 
\begin{equation}
d=\int{d^2\tau d^2\tau_1\over 2 \pi^2}\,{1\over \tau_1^2
\tau_2^2}\, [N(\tau_1)+N(\tau_2)-N(\tau)-N(\tau_1)N(\tau_2)]
\label{reas3}
\end{equation}
can only be found once the scaling solution $N(\tau)$ is known.

For  a fixed coupling constant, the saturation scale grows exponentially with rapidity
\begin{equation}
Q_s^2(Y)=Q_0^2\,\exp{\left[\Delta Y\right]},
\label{qsyfix}
\end{equation}
where $\bar \alpha_s=\bar \alpha_0=$ constant, $\Delta=d\bar \alpha_0$ and 
$Q^2_0=Q^2_s(Y=0)$ (i.e., the evolution starts at $Y=0$).

For running coupling, the momentum scale is expected to be $\sim Q_s(Y)$, 
suggesting% 
\footnote{This approximation is also supported by numerical results 
\cite{Braun:2000wr,Armesto:2001fa,Golec-Biernat:2001if} 
which show
that in momentum space the typical transverse momentum of the gluons 
is $\sim Q_s$.} the substitution
$\bar \alpha_s\to \bar \alpha_s(Q_s(Y))$ in Equation (\ref{qsy}).
This leads to \cite{Iancu:2002tr}
\begin{equation}
Q_s^2(Y)=\Lambda^2 \,\exp{\left[\Delta^\prime\sqrt{Y+X}\right]},
\label{qsyrun}
\end{equation}
where $(\Delta^\prime)^2=24N_c d/\beta_0$
and $X=(\Delta^\prime)^{-2}\ln{(Q_0^2/\Lambda^2)}$. 

\begin{figure}[h!]
\begin{center}
\epsfig{file=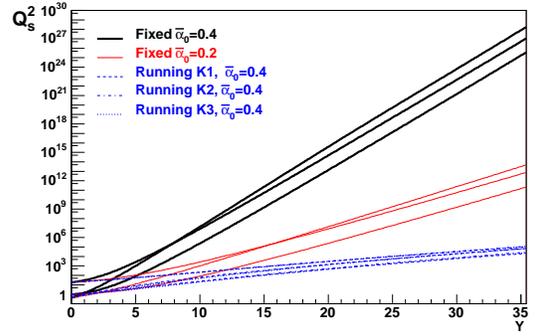,width=0.40\textwidth}
\end{center}
%\vskip -1.cm
\caption{The rapidity dependence of the saturation momentum $Q_s^2$ 
for fixed $\bar \alpha_s=0.4$ (thick solid),
fixed $\bar \alpha_s=0.2$ (thin solid), and running coupling with $\bar
\alpha_0=0.4$ for kernels $K1$ (dashed), $K2$ (dashed-dotted) and $K3$ (dotted
lines).  For each group, lines from top to bottom in the rightmost side
correspond to initial conditions AS with $c=1.17$, MV and GBW.}
\label{fig:qsat}
\end{figure}

The $Y$-dependence of $Q_s^2$ for several initial conditions and for different choices of $\bar \alpha_0$, calculated for all the kernels considered in this work, is shown in Figure \ref{fig:qsat}.

The rise of $Q_s^2$ is much faster for fixed than running coupling, in accordance with (\ref{qsyfix}) and 
(\ref{qsyrun}) and as already observed in \cite{Lublinsky:2001yi,
Golec-Biernat:2001if,
Triantafyllopoulos:2002nz,Braun:2003un,Rummukainen:2003ns,Kutak:2004ym,
Khoze:2004hx,Chachamis:2004ab}

For fixed coupling constant, $Q_s^2$ exhibits with good accuracy
an exponential
behaviour for high-enough values of $Y$. The value of the slope 
extracted from a fit to the function (\ref{qsyfix}) is $\Delta\simeq 1.83$
for $\bar \alpha_0=0.4$. As expected, for $\bar \alpha_0=0.2$ this value 
is reduced by a factor two, $\Delta\simeq 0.91$. For the constant 
(\ref{reas3}), we find $d\simeq 4.57$, in agreement
with previous numerical studies at very high rapidities \cite{Albacete:2003iq}
but slightly smaller than the theoretical expectation $d=4.88$
\cite{Iancu:2002tr,Mueller:2002zm}.
%In previous numerical studies 
%\cite{Armesto:2001fa,Levin:2001et,Golec-Biernat:2001if},
%an even smaller value of $d \sim 4.1$ was obtained. We have checked that 
%this is due to the fact that the rapidity region for the fit in our case 
%corresponds to much larger $Y$.

For the case of a running coupling constant, an exponential fit 
can be done only for a very limited $Y$-region. For example, for 
$Y\sim 10$ we find a logarithmic slope $\sim0.28$ for GBW or MV initial
conditions with $Q_0 \sim 1$ GeV, in
agreement with the results of \cite{Triantafyllopoulos:2002nz} but
smaller than the values found in
\cite{Khoze:2004hx} (see also
\cite{Kutak:2004ym,Chachamis:2004ab}).
The exponential function (\ref{qsyfix}) is unable to fit the full
$Y$-range. In contrast, the weaker rapidity dependence of (\ref{qsyrun}) 
does provide a good fit in the full $Y$-range.
The fit to (\ref{qsyrun})  
yields $\Delta^\prime \simeq 3.2$, while the theoretical expectation
\cite{Iancu:2002tr,Mueller:2002zm} is slightly larger, $\Delta^\prime=3.6$.
%We finally note that in \cite{Rummukainen:2003ns} the $Y$-derivative of 
%$\ln{Q_s^2(Y)}$ has been
%found numerically to be
%proportional to $\sqrt{\alpha_s(Q_s(Y))}$ in a much more
%restricted range of $Y$. We have been unable to fit our results over 
%the full $Y$-range to the corresponding $Y$-dependence, $Q_s^2(Y)\propto
%\exp{Y^{2/3}}$.

\subsection{Nuclear size dependence of the saturation scale}

\begin{figure}[h!]
\begin{center}
\epsfig{file=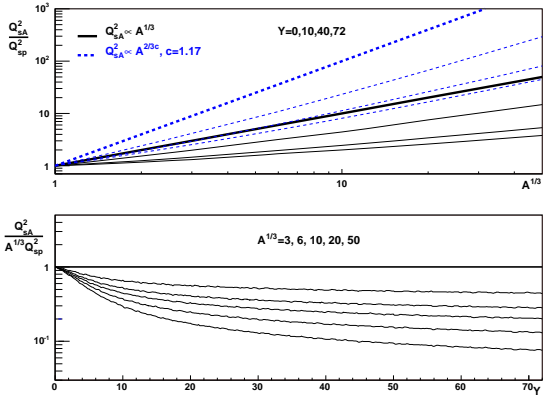,width=0.45\textwidth}
\end{center}
%\vskip -1.cm
\caption{Upper plot: $Q_{sA}^2/Q_{sp}^2$ versus $A^{1/3}$ for initial
conditions GBW ($Q_{sA}^2(Y=0)\propto A^{1/3}$, solid) and AS with $c=1.17$
($Q_{sA}^2(Y=0)\propto A^{2/3c}$, dashed lines); thick lines are the
results for $Y=0$ in the running coupling case and for all rapidities in fixed
coupling; for running coupling, different rapidities $Y=10$, 40 and 72 (thin
lines) are
shown from top to bottom for each initial condition. Lower plot:
$Q_{sA}^2/(A^{1/3}Q_{sp}^2)$ versus $Y$ for GBW with
$A^{1/3}=3$, 6, 10, 20 and 50 with
the same line convention as the upper plot (the results for fixed coupling
have been obtained for $Y<36$ and extrapolated as a constant equal to 1).
In all plots
$\bar \alpha_0=0.4$ and in the running coupling case the kernel $K1$ has been
used.}
\label{fig:adep}
\end{figure}

The nuclear size enters the initial condition. Here, we examine the effect of BK evolution on the initial nuclear size dependence of the saturation scale.
Figure \ref{fig:adep} summarises our results. 

In the fixed coupling case, the initial $A$-dependence of the saturation scale is preserved, irrespectively  of whether this dependence is $\propto A^{1/3}$ 
as for the GBW or MV initial conditions (which produces numerical results for
the $A$-dependence which are very close
to those obtained for GBW), or whether it
differs from $\propto A^{1/3}$ 
due to an anomalous dimension as the one included in the AS initial condition. 
This is in agreement with the theoretical expectation that, given the dilatation invariance of the BK equation (\ref{eq:bk}), any nuclear dependence present in the initial condition can be scaled out.

The evolution with running coupling is seen to reduce the $A$-dependence with increasing rapidity.
If fitted in a wide rapidity range, the dependence of $\ln{[Q_{sA}^2(Y)/ Q_{sp}^2(Y)]}$ on $Y$ is $\sim Y^{-0.4}$. However,  for large values of
$A$ and $Y$, the decrease with increasing $Y$ is $\propto 1/\sqrt{Y}$ and 
thus well described by  \cite{Mueller:2003bz}
\begin{equation}
\ln{{Q_{sA}^2(Y)\over Q_{sp}^2(Y)}}\simeq {\ln^2{\left[
{h\,Q_s^2(Y=0)\over \Lambda^2}\right]}\over 2\sqrt{(\Delta^\prime)^2Y}}\,, 
\label{finmue}
\end{equation}
where $h\,Q_s^2(Y=0)$ is the initial saturation momentum for the nucleus, 
and $(\Delta^\prime)^2$ is defined below Equation (\ref{qsyrun}).

\section{Geometric scaling and experimental data}

%In this section we will look for the presence of the properties of the
%solutions of the BK equations in experimental data, in particular the
%existence of a scaling solution, which leads to the phenomenon of 
%{\it geometric scaling} and the behaviour of the saturation scale with found in lepton-proton data
%these solutions is the existence of a scaling cuve at asymptotic energies.
%This {\it geometric scaling} was first found in lepton-proton data
%\cite{Stasto:2000er} and extended to other observables in
%\cite{Armesto:2004ud}. 

\subsection{Geometric scaling in lepton-hadron collisions}

Following \cite{Armesto:2004ud} and motivated by the previous sections, let us
assume that both the energy and the nuclear size (or centrality) 
dependence of the scattering amplitude $N({\bf r},x;{\bf b})$ 
can be encoded in the saturation scale $Q_{s, h}
(x,{\bf b})$ for any $h$ (proton or nucleus). Then, the cross section
(\ref{eq1}) can be written as
\begin{eqnarray}
  \sigma_{T,L}^{\gamma^* h}(x,Q^2) &=&  \pi R_h^2
 \int d{\bf r} \hspace{-0.1cm}
  \int_0^1
  \hspace{-0.2cm} dz
  \vert \Psi_{T,L}^{\gamma^*}(Q^2,{\bf r},z)\vert^2 \nonumber \\
  &\times& 2 \int d{\bf \bar b}\, N_h(r\qsh(x,{\bf \bar b}))\, .
\label{eqnuclsc}
\end{eqnarray}
In this case, since $\vert \Psi_{T,L}^{\gamma^*}(Q^2,{\bf r},z)\vert^2$ is
proportional to $Q^2$ times a function of ${\bf r}^2 Q^2$, the cross section
is only a function of $\tau_h^2= Q^2 / \qsh^2(x)$. This {\it geometric
scaling} was found to describe all lepton-proton data with $x<0.01$
\cite{Stasto:2000er}. 
In order to compare  protons and different nuclei \cite{Armesto:2004ud} 
one needs to make some assumptions about the
impact parameter dependence. Here, we have assumed
that all the $b$ dependence can 
be scaled by the nuclear radius of the hadronic target%
\footnote{This is exact for trivial impact parameter dependences as a
step-function or a gaussian. We have checked that it gives a rather good
approximation for a realistic, Wood-Saxon, profile.} $h$, 
${\bf \bar b}={\bf b}/\sqrt{\pi R_h^2}$, with $R_A=(1.12 A^{1/3}-0.86 A^{-1/3})$
fm. 
%(This will work for trivial impact
%parameter dependences as a step-function or a gaussian, but would need of some
%corrections for more realistic profile functions).
%For a trivial impact parameter dependence of the saturation
%scale, $\qsh(x,{\bf b}) = \qsh(x)\, \Theta(R_h - b)$, and 
%Since $\vert \Psi_{T,L}^{\gamma^*}(Q^2,{\bf r},z)\vert^2$ is
%proportional to $Q^2$ times a function of ${\bf r}^2 Q^2$, 
%Eq. (\ref{eqnuclsc}) depends solely on $\tau_h^2= Q^2 / \qsh^2(x)$ and the
Then, the condition for geometric scaling in lepton-nucleus data is
\begin{equation}
  \frac{\sigma^{\gamma^*A}(\tau_A)}{\pi R_A^2}=
  \frac{\sigma^{\gamma^*p}(\tau_A)}{\pi R_p^2}\, .
  \label{eqnormal}
\end{equation}
For the $A$-dependence, we make the ansatz that the saturation
scale in the nucleus grows with the ratio of the transverse
parton densities to some power $1/\delta$, which we take as a free parameter
\begin{equation}
   \label{eqtaua}
   \qsa^2=\qs^2\left(\frac{A \pi R_p^2}
   { \pi R_A^2}\right)^\frac{1}{\delta} \hspace{-0.1cm}
   \Rightarrow 
   \tau_A^2=\tau_p^2\left(\frac{ \pi R_A^2}{A 
                       \pi R_p^2}\right)^\frac{1}{\delta} .
\end{equation}
Here, $\pi R_p^2$ is the second free parameter to be fixed by the data.
 
\begin{figure}[hbt]
\epsfxsize=6cm
%\vskip -.2cm
\centerline{\epsfbox{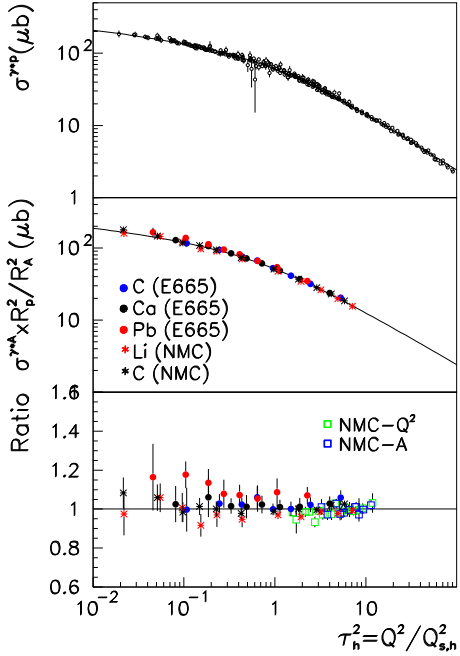}}
%\vskip -0.3cm
\caption{Geometric scaling for $\gamma^*p$ (upper panel, data
from~\cite{proton}),
$\gamma^*A$ (middle panel, data from \cite{Adams:1995is,Arneodo:1995cs}) 
and the ratio of data for $\gamma^*A$ over the prediction
from (\ref{eqscalf}) (lower panel). As an additional check,
the lower plot also shows data for $\gamma^*A$  
normalized with respect to $\gamma^*C$ \cite{Arneodo:1996rv}, and divided by
the corresponding prediction from Eq. (\ref{eqscalf}).}
\label{figprot}
\end{figure}

For the proton case we take the Golec-Biernat and W\"usthoff (GBW)
parametrization~\cite{Golec-Biernat:1998js}, $\qs^2=(\bar x/x_0)^{-\lambda}$
in GeV$^2$, $x_0= 3.04\cdot 10^{-4}$ and $\lambda=0.288$. This parametrization
shows geometric scaling as can be seen in Fig.~\ref{figprot}.
In order to proceed to the nuclear case, we need a functional form of
the scaling curve. 
The data~\cite{proton} are seen to be well parametrized  by \cite{Armesto:2004ud}
\begin{equation}
  \sigma^{\gamma^* p}(x,Q^2) \equiv
  \Phi(\tau_p^2) = 
\bar\sigma_0
  \left[ \gamma_E + \Gamma\left(0,\xi\right) +
         \ln\xi \right]\, ,
       \label{eqscalf}
\end{equation}
where $\gamma_E$ is the Euler constant, $\Gamma\left(0,\xi\right)$
the incomplete $\Gamma$ function and $\xi=a/\tau_p^{2b}$,
with $a=1.868$ and $b=0.746$. The normalization is fixed by
$\bar\sigma_0=40.56$ mb. 
%For our purpose, Eq.~(\ref{eqscalf}) 
%is just a convenient ansatz for the scaling function $\Phi(\tau_h)$. 

To determine $\qsa^2$, we use Eqs. (\ref{eqnormal}) and (\ref{eqtaua}) and
compare the functional shape 
(\ref{eqscalf}) to the available experimental data for $\gamma^*A$ 
collisions with $x\leq 0.0175$ 
\cite{Adams:1995is,Arneodo:1995cs,Arneodo:1996rv}, using 
$\xi = a/\tau_A^{2b}$.
%The parameters $\delta$ and $\pi R_p^2$ in 
%(\ref{eqnormal}) -- (\ref{eqscalf}) are fitted by $\chi^2$ 
%minimization adding the statistical 
%and systematic errors in quadrature. 
We obtain $\delta=0.79\pm0.02$ and $\pi R_p^2=1.55 \pm 0.02$
fm$^2$ for a $\chi^2/{\rm dof} = 0.95$  (see Fig.~\ref{figprot} for the
comparison). Notice that these parameters translate into a growth of the
saturation scale {\it faster} than $A^{1/3}$ for large nuclei. 
If we impose $\qsa^2\sim A^{1/3}$ in the fit, a much worse value of
$\chi^2/{\rm dof} =2.35$ is obtained. We conclude that the small-$x$
experimental data on $\gamma^*A$ collisions favours an increase of 
$\qsa^2$ faster than $A^{1/3}$. The numerical coincidence 
$b\simeq \delta$ is consistent with the absence of shadowing
in nuclear parton distributions at $Q^2\gg \qsa^2$.

%%%%%%%%%%%%%%%%%%%%%%%%%%%%%%%%%%%%%%%%%%%%%%%%%%%%%%%%%%%%%%%%%%%%
%%%%%%%%%%%%%%%%%%%%%%%%%%%%%%%%%%%%%%%%%%%%%%%%%%%%%%%%%%%%%%%%%%%%%
%\vskip -.2cm

\subsection{Geometric scaling and multiplicities in AA collisions}

In \cite{Armesto:2004ud} a simple formula for multiplicities in symmetric
colliding systems at central rapidities has been proposed:

\begin{eqnarray}
\frac{1}{N_{\rm part}}
\frac{dN^{AA}}{d\eta}\Bigg\vert_{\eta\sim 0}=N_0\sqrt{s}^\lambda
N_{\rm part}^{\frac{1-\delta}{3\delta}}\, .
\label{eqmult}
\end{eqnarray}
An easy way to derive this formula is to take,
as the starting point, the factorized
expression~\cite{Gribov:tu} for gluon production 
\begin{equation}
   \frac{dN^{AB}_g}{dyd{\bf p}_t d{\bf b}}\propto \frac{\alpha_S}{{\bf p}_t^2}
   \int d{\bf k}\  \phi_A(y,{\bf k}^2,{\bf b})\,
    \phi_B\left(y,({\bf k}-{\bf p}_t)^2,{\bf b}\right)\, ,
   \label{eqfact}
\end{equation}
where $\phi_h(y,{\bf k},{\bf b})=\int d{\bf r}\,\exp\{i{\bf r\cdot k}\}\,
N_h({\bf r},x;{\bf b})/(2\pi r^2)$~\cite{facto}. 
%$y = \ln 1/x $. 

\begin{figure}
\epsfxsize=5.5cm
\centerline{\epsfbox{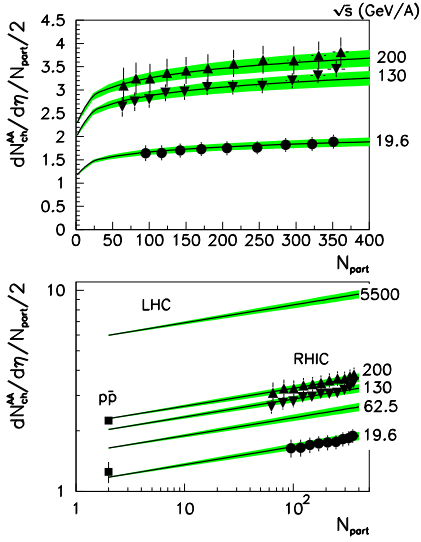}}
%\vskip -0.3cm
\caption{Energy and centrality dependence of the multiplicity of charged 
particles in Au+Au collisions (\ref{eqmult}) compared to PHOBOS data
\cite{Back:2002uc}.
Also shown in the lower panel are the ${\bar p}$+$p$ data 
\cite{protmult} and results for $\sqrt{s}=$ 62.5 and 5500 GeV/A.}
\label{figmult}
\end{figure}

If one assumes geometric scaling for the parton distributions, 
$\phi_A(y,{\bf k}^2,{\bf b})$$\equiv$
$\phi({\bf k}^2/\qsa^2(y,{\bf b}))$, we obtain
\begin{eqnarray}
  &&\frac{dN^{AA}_g}{dy}\Bigg\vert_{y\sim 0}\propto 
  \int \frac{d{\bf p}_t}{{\bf p}_t^2} d{\bf k} d{\bf b}\ 
  \phi\left(\frac{{\bf k}^2}{\qsa^2}\right)\,
  \phi\left(\frac{({\bf k}-{\bf p}_t)^2}{\qsa^2}\right)\nonumber \\
  &&=\qsa^2 \pi R_A^2\int \frac{d{\bf s}}{{\bf s}^2} d{\bf \tau} 
  d{\bar {\bf b}}\ 
  \phi({\bf \tau}^2)\,\phi\left( ({\bf \tau}-{\bf s})^2\right),
  \label{eqsym}
\end{eqnarray}
where the integrand is a constant. This proportionality between the total
multiplicities and the saturation scale is shared by other models of
hadroproduction~\cite{Gribov:tu,facto,Eskola:1999fc,Kovchegov:2000hz}.
It is important to notice, however, that for the case of geometric scaling, 
Eq.~(\ref{eqmult}) is more general than the factorized form (\ref{eqfact}).
Indeed, any functional shape of the integrand will lead to the same result
providing geometric scaling holds. In order to recover Eq.~(\ref{eqmult}), 
the energy dependence of the saturation scale in (\ref{eqsym}) is given by the 
GBW parameter $\lambda=0.288$; for its centrality dependence 
we use the known proportionality in symmetric
A+A collisions between 
the number $N_{\rm part}$ of participant nucleons 
and the nuclear size $A$ and 
$\qsa^2 \propto A^{1/3\delta}$, with $\delta=0.79\pm 0.02$. 
In this way, the energy and centrality
dependences are determined by parameters fitted to $\gamma^*-p$ and $\gamma^*-A$
respectively. 
In all these models, the hadron yield is assumed to be proportional
to the yield of produced partons. 
The remaining normalization constant is fixed to $N_0=0.47$ in
order to reproduce PHOBOS data, see Fig.~\ref{figmult}. It is interesting that
even the ${\bar p}$+$p$ data (\cite{protmult}, as quoted
in \cite{Back:2002uc}) at $\sqrt{s}=$ 19.6 and 200 GeV are accounted for.
Eq.~(\ref{eqmult}) implies that the energy and the centrality dependence of the 
multiplicity factorize, in agreement with the results by 
PHOBOS~\cite{Back:2002uc}.
%
%%%%%%%%%%%%%%%%%%%%%%%%%%%%%%%%%%%%%%%%%%%%%%%%%%%%%%%%%%%%%%%%%%%%%%%%%
%%%%%%%%%%%%%%%%%%%%%%%%%%%%%%%%%%%%%%%%%%%%%%%%%%%%%%%%%%%%%%%%%%%%%%%%%

A formula similar to Eq.~(\ref{eqmult}) has been extensively employed to study
multiplicities in Au+Au collisions in \cite{kln}. These authors assume
$\qsa^2\propto A^{1/3}$ and an additional enhancement factor
$\ln(\sqrt{s}^\lambda N_{\rm part}^{1/3})$
argued to come from scaling violations of the coupling constant in 
(\ref{eqfact}). Notice that
for the accessible
range of $A$, $A^{4/9} \sim A^{1/3}\, \ln(A^{1/3})$ -- this being the
reason why both approaches provide a fair description of the data at RHIC.

\subsection{Geometric scaling and dAu data}

\begin{figure}
\epsfxsize=6cm
\centerline{\epsfbox{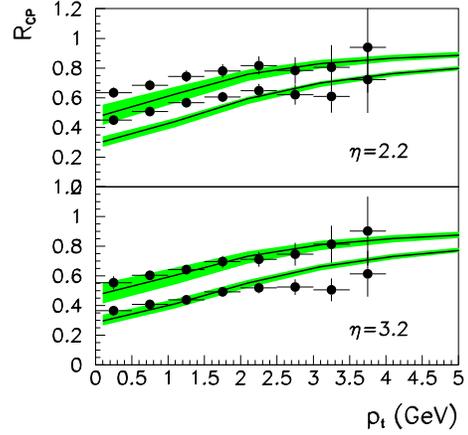}}
%\vskip -0.3cm
\caption{Normalized ratios of central and semi-central to peripheral d+Au 
collisions measured by BRAHMS \cite{Arsene:2004ux} compared to results from
Eq.~(\ref{eqratpt}). The bands represent the uncertainty in the determination
of $N_{\rm coll}$~\cite{Arsene:2004ux}.}
\label{figbrahms}
\end{figure}

The forward rapidity region of the RHIC experiments has become the main 
testing ground for saturation ideas. In this section, we check to
which extent geometric scaling can account for the observed
suppression on particle yields. The situation in this case is, however, more
model-dependent than in the previous two cases. The reason is the existence of 
two different saturation scales (one for
the deuteron and another one for the gold nucleus) which precludes 
the trivial changes of variables done before. In order to proceed, we
observe that if one writes Eq.~(\ref{eq1}) in momentum space $k$, the main
contribution to the cross section comes from $k\sim Q/2$. Approximating the
dipole wave function (in momentum space) by $\delta(k-Q/2)$,
the scaling curve (\ref{eqscalf}) is (except for
a normalization constant) the unintegrated gluon distribution, 
$\phi_A(k=Q/2) \simeq \Phi(\tau_A)$, with $\tau_A=k^2/4\bar\qsa^2$. For the
case of particle production in dAu collisions, the gluon saturation scale 
$\bar\qsa^2= N_c\qsa^2/C_F$ needs to be used. Now, if the parton distribution
in the deuteron falls off sufficiently quickly,
$\phi_d\sim 1/k_t^n$, $n\gg 1$, we obtain from Eq. (\ref{eqfact})
\begin{equation}
   \frac{\frac{dN^{\rm dAu}_{c_1}}{N_{\rm coll_1}d\eta d^2p_t}}
   {\frac{dN^{\rm dAu}_{c_2}  }{N_{\rm coll_2}d\eta d^2p_t}}
   \approx
   \frac{N_{\rm coll_2}\phi_A(p_t/Q_{\rm s,c_1})}
    {N_{\rm coll_1}\phi_A(p_t/Q_{\rm s,c_2})}
   \approx
   \frac{N_{\rm coll_2}\Phi(\tau_{c_1})}{N_{\rm coll_1}\Phi(\tau_{c_2})}\, 
   \label{eqratpt}
\end{equation}
for the centrality classes $c_1$, $c_2$. This simple formula relates the
suppression measured in lepton-nucleus collisions to that in d+Au.
For the comparison in Fig.~\ref{figbrahms} to data~\cite{Arsene:2004ux} 
on the normalized yields of central and semi-central over peripheral 
d+Au collisions, we use the number of nucleon-nucleon 
collisions $N_{\rm coll}$ in different 
centrality bins~\cite{Arsene:2004ux} with $N_{\rm coll_1} =
13.6\pm0.3$, $7.9\pm0.4$ and $N_{\rm coll_2} =
3.3\pm0.4$. Only the two most forward rapidities $\eta=2.2$ and 3.2 
are compared. This simplistic analysis indicates that
the suppression of particle yields in forward rapidity measured in d+Au
collisions is in agreement, through geometric scaling, with the nuclear 
shadowing measured in lepton-nucleus collisions. The connection between the 
small $x$- and $A$-dependence of 
parton distribution functions, and the suppression of normalized 
yields in d+Au collisions~\cite{Arsene:2004ux} at forward rapidity has 
been discussed in several recent 
works~\cite{Albacete:2003iq,others}.
Eq.~(\ref{eqratpt}) contributes to
this discussion by illustrating to what extent the suppression of 
high-$p_t$ particles in d+Au at RHIC can be accounted for by the 
shadowing in $\gamma^*A$ collisions, see Eq.~(\ref{eqscalf}).

%
%%%%%%%%%%%%%%%%%%%%%%%%%%%%%%%%%%%%%%%%%%%%%%%%%%%%%%%%%%%%%%%%%%%%%%%%
%%%%%%%%%%%%%%%%%%%%%%%%%%%%%%%%%%%%%%%%%%%%%%%%%%%%%%%%%%%%%%%%%%%%%%%%%

\section{Comparison of data and theory and conclusions}

The two main properties of the BK evolution are the existence of a scale which
indicates the presence of a saturated region and the existence of a scaling
solution at asymptotic energies. The analytic form of the asymptotic solution
is only known for some ranges of the dipole size. Here we have presented a
numerical analysis of the BK equations without impact parameter dependence. In
order to study the possible effects of next-to-leading-log corrections to the
BK equation we have included 
the running of the coupling constant, using several prescriptions. Our results
confirm the existence of a scaling solution also for the running coupling
case. Most of our numerical results agree with the analytical 
estimations. Namely,  the evolution in rapidity of the saturation scale has
been found to be exponential for the case of fixed coupling and slower (the
exponential of a square root) for the running coupling case and with
parameters in agreement with analytical estimates; the nuclear size
dependence of the saturation scale has been found to be preserved in the fixed
coupling case and to disappear asymptotically in the running coupling case;
the small-$r$ behaviour of the scaling solutions reproduces the {\it anomalous
dimension} found analytically for the case of a fixed coupling, but it is
found to be different (in disagreement with analytical estimations) in the
running coupling case. 

If the dynamics of the BK equations is relevant for the present experimentally 
accessible regimes, the above properties should be realized in the data. The
observation~\cite{Stasto:2000er}
that all the available data on lepton-proton scattering with $x < 0.01$
can be described by a single variable $\tau^2=Q^2/\qs^2$ is reminiscent
of the existence of a scaling solution. The energy dependence of the
saturation scale extracted from a fit to experimental
data~\cite{Golec-Biernat:1998js} is too small compared with the results of BK
with fixed coupling. It is, however, in approximate agreement with the running
coupling case if one is restricted to a small range of energies. In the
nuclear case, we have found~\cite{Armesto:2004ud} that a similar geometric
scaling of the nuclear data is possible by encoding the nuclear size in the
saturation scale. The $A$-dependence is, however, different from the usually
assumed $A^{1/3}$ and turns out to be stronger,  $\sim A^{4/9}$. If this
result is not a numerical accident, this behaviour could only have a dynamical
origin (the geometrical behaviour of $\qsa^2$ is $A^{1/3}$). This is not in
disagreement with the solutions we have obtained from the BK equations, but
the relation is obscure: the $A$-dependence of the saturation scale is
preserved (i.e. it is the same as in the initial conditions) for the fixed
coupling case, but disappears for the running coupling case. The nuclear size
dependence is, however, expected to be modified when an appropriate treatment
of the impact parameter is taken into account in the BK equation
\cite{bdeprc}. 

We have also found that a
simple explanation for the centrality dependence of the multiplicities
measured in symmetric systems at central rapidities is possible
with the nuclear size dependence of the saturation scale obtained from
lepton-nucleus data. In the presence of a geometric scaling, these
multiplicities are proportional to $\qsa^2$. This translates into a
proportionality of the multiplicities with the number of participants for the
geometrical estimation $\qsa^2\sim A^{1/3}$. The $A$-dependence $\qsa^2\sim
A^{4/9}$ gives, however, a nice description of the experimental data. As a
final check, we have found a striking similarity of the suppression of nuclear
yields in forward d+Au collisions measured by the BRAHMS collaboration at
RHIC and the nuclear shadowing measured in lepton-nucleus experiments. All
these observations are in agreement with the interpretation of the data in
terms of saturation physics, but the existence of a numerical accident cannot
be excluded. A quantitative description of the experimental data with
explicit use of QCD evolution equations will be needed in order to 
establish the relevance of the saturation effects.

\begin{acknowledgement}
J.~L.~A. and J.~G.~M.  acknowledge financial support
by the Ministerio de Educaci\'on y Ciencia of Spain (grant no. AP2001-3333)
and the Funda\c c\~ao para a Ci\^encia e a Tecnologia  of Portugal (contract
SFRH/BPD/12112/2003) respectively.
\end{acknowledgement}

\end{document}